\documentclass[preprints,article,accept,pdftex,moreauthors]{Definitions/mdpi} 

\firstpage{1} 
\pubvolume{1}
\issuenum{1}
\articlenumber{0}
\pubyear{2024}
\copyrightyear{2024}
\datereceived{ } 
\daterevised{ } 
\dateaccepted{ } 
\datepublished{ } 
\hreflink{https://doi.org/} 

\Title{Gravitational Waves: \\
\large{Echoes of the Biggest Bangs since the Big Bang and/or BSM Physics?}}

\TitleCitation{Gravitational Waves}

\Author{John Ellis $^{1}$,*\orcidA{0000-0002-7399-0813}}

\AuthorNames{John Ellis}

\AuthorCitation{Ellis, J.}

\address{%
$^{1}$ \quad Physics Department, King's College London, Strand, London WC2R 2LS, United Kingdom}

\corres{John.Ellis@cern.ch}

\abstract{\\
{\it ``If one could ever prove the existence of gravitational waves, the processes responsible for their generation would probably be much more curious and interesting than even the waves themselves.” (Gustav Mie, 1868 - 1957)}\\
The discovery of gravitational waves has opened new windows on astrophysics, cosmology and physics beyond the Standard Model (BSM). Measurements by the LIGO, Virgo and KAGRA Collaborations of stellar-mass binaries and neutron star mergers have shown that gravitational waves travel at close to the velocity of light, and also constrain BSM possibilities such as a graviton mass and Lorentz violation in gravitational wave propagation. Follow-up measurements of neutron star mergers have provided evidence for the production of heavy elements, possibly including some essential for human life. The gravitational waves in the nanoHz range observed by Pulsar Timing Arrays (PTAs) may have been emitted by supermassive black hole binaries, but might also have originated from BSM cosmological scenarios such as cosmic strings, or phase transitions in the early Universe. The answer to the question in the title may be provided by gravitational-wave detectors at higher frequencies, such as LISA and atom interferometers.\\
~~\\
KCL-PH-TH/2024-05}


\begin{document}




\section{Prediction of gravitational waves}

This article is intended as an introduction to the phenomenology and interpretation of gravitational waves (GWs) that might be of interest to those, like myself, who are neophytes in the field. Readers of this article will already know that Albert Einstein predicted the existence of GWs in 2016 but this statement is a simplification, like so many about the convoluted history of physics.~\footnote{For informative summaries, see~\cite{Cervantes-Cota:2016zjc,rothman2018secret}.} For starters, many physicists had previously conjectured that GWs should exist, though they did not have the correct theory of gravity that was later provided by Einstein. In particular, in 1893 Oliver Heaviside predicted the existence of GWs in a field theory of gravity that he had formulated~\cite{heaviside1893gravitational} and in 1905 Henri Poincar{\'e} also conjectured the existence of GWs~\cite{poincare1905dynamique}. Secondly, Einstein initially thought that his General Theory of Relativity did {\it not} predict the existence of GWs, writing to Schwarzschild in early 1916 that {\it ``“There are no gravitational waves analogous to light waves”}. Then, when Einstein first predicted GWs in a paper written later in 1916, he predicted erroneously the existence of {\it monopole} radiation~\cite{einstein1916}. He acknowledged his error in 1918, predicting correctly that GWs should be {\it tensorial} and giving the correct radiation formula~\cite{einstein1918gravitationswellen}. However, Einstein was always sceptical about the reality of GWs. For example, he and Nathan Rosen wrote a paper in 1936 that denied their existence. However, they were quickly corrected by Howard Robinson, who pointed out that they had made a poor choice of coordinates, and the published version of the Einstein-Rosen paper confirms the prediction of GWs~\cite{Einstein:1937qu}. Nevertheless, Einstein’s scepticism rubbed off on the broader physics community, which was not completely convinced about the reality of GWs until the late 1950s, following the work of Felix Pirani~\cite{pirani1956physical}, Hermann Bondi, Ivor Robinson and Andrzej Trautmann: see~\cite{Cervantes-Cota:2016zjc,rothman2018secret} for historical reviews.

\section{Detection of gravitational waves}

The first, indirect, evidence for GWs was provided by measurements of the celebrated binary pulsar PSR B1913+16 that was discovered by Hulse and Taylor in 1974~\cite{Hulse:1974eb}. Measurements of the rate of change of the periodicity of electromagnetic pulses from this system over four decades are highly consistent with predictions based on the emission of GWs, which causes energy loss and inspiral of the binary components. The measured rate of change is $0.9983 \pm 0.0016$ of the predicted value~\cite{Weisberg:2016jye}, agreement at the level of $1.6 \times 10^{-3}$. However, measurements of the binary pulsar system PSR J0737-3039 A,B are even more precise, agreeing with the value predicted from GW emission in general relativity at the level of $0.999963\pm0.000063$~\cite{Kramer:2021jcw}, verifying the theoretical prediction at the level of $6 \times 10^{-5}$.

\begin{figure}
    \centering
    \includegraphics[width=0.8\textwidth]{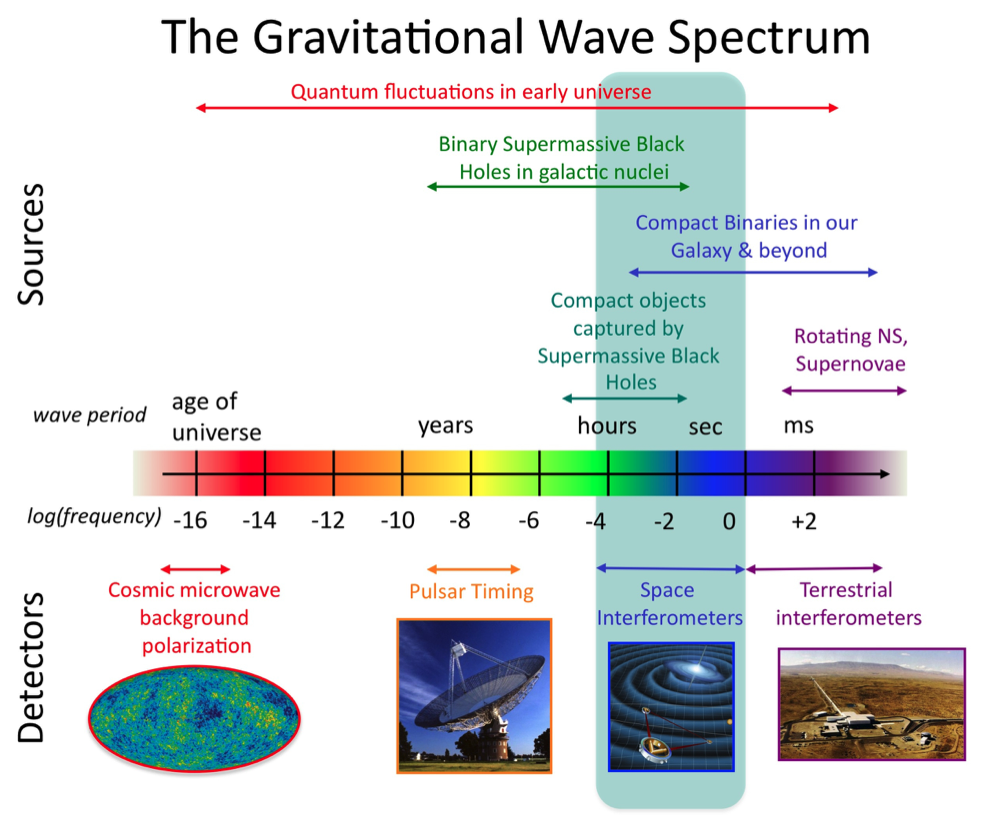}
    \caption{The gravitational wave spectrum as a function of wavelength and frequency together with some possible astrophysical and cosmological sources and some detection techniques. Figure credit: NASA Goddard Space Flight Center.}
    \label{fig:GWspectrum}
\end{figure}

The direct discovery by the LIGO and Virgo laser interferometer experiments of GWs emitted by mergers of stellar-mass black hole (BH) binaries~\cite{LIGOScientific:2016aoc} opened new windows on gravitational physics, astrophysics and cosmology, as well as potential opportunities to explore beyond the Standard Model (BSM) physics, as illustrated in Fig.~\ref{fig:GWspectrum}. These new windows are the main theme of this article.

Before the discovery of GWs, astrophysical calculations had indicated a lower bound on BH masses from stellar collapses of ${\cal O}(5)$ solar masses, and the existence of BHs with masses just above this limit had been inferred from observations of binary systems emitting $X$ rays. Other calculations had indicated that there should be a gap in BH masses from stellar collapses above ${\cal O}(60)$ solar masses, where an $e^+ e^-$ pair-production stability would set in~\cite{Woosley:2021xba}. The first LIGO-Virgo mergers involved BHs with masses comfortably between these limits, though somewhat heavier than the $X$-ray binaries, but some mergers observed subsequently featured objects lighter than the lower limit, and others featured objects above the upper limit. Much astrophysical interest centres on the objects outside these limits, particularly heavier objects.

\begin{figure}
    \centering
    \includegraphics[width=0.8\textwidth]{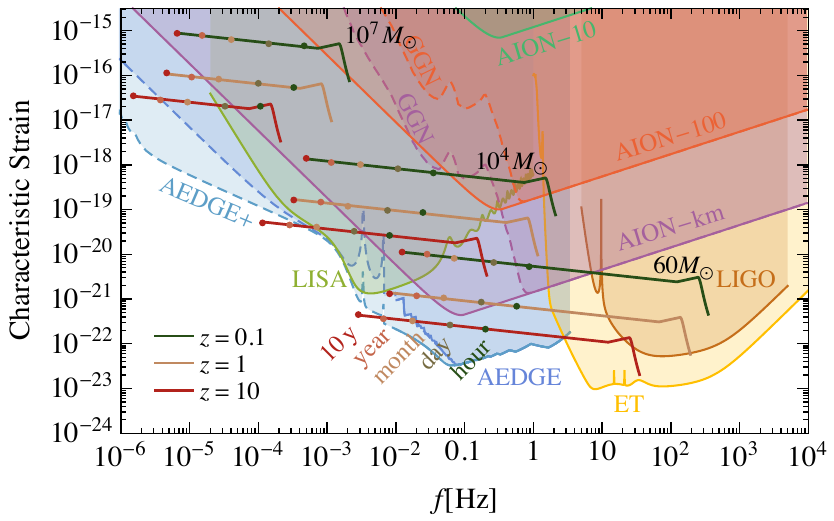}
    \caption{Strain sensitivities of LIGO, LISA, ET and the proposed atom interferometers AION-10, -100, -km, and AEDGE, compared with the signals expected from mergers of equal-mass binaries whose masses are 60, $10^4$ and $10^7$ solar masses. The assumed redshifts are $z = 0.1, 1$ and 10, as indicated. The dots indicate the times remaining during the inspiral stage before the final mergers~\cite{Badurina:2021rgt}.}
    \label{fig:strains}
\end{figure}

Fig.~\ref{fig:strains} illustrates the sensitivities to GW strains of LIGO and Virgo~\cite{LIGOScientific:2016aoc}, which are optimised around $10^2$~Hz, of the proposed follow-on terrestrial laser interferometer Einstein Telescope (ET)~\cite{Maggiore:2019uih}, of the planned space-borne laser interferometer LISA~\cite{Barausse:2020rsu,Colpi:2024xhw}, which is optimised around $10^{-2}$~Hz, and of prospective atom interferometers (AIs), AION~\cite{Badurina:2019hst} and AEDGE~\cite{AEDGE:2019nxb} (see Section~7), which are optimised around $0.1$ to 1~Hz,. Also shown are the signals expected from mergers of equal-mass binaries whose masses are 60, $10^4$ and $10^7$ solar masses at various redshifts. 

\section{Looking for BSM physics}

The measurements of GWs from BH mergers can be used directly to constrain possible modifications of general relativity and search for BSM physics. For example, LIGO-Virgo measurements constrain the graviton mass, which would alter GW velocity as a function of frequency and hence distort the characteristic chirp pattern: $m_g < 1.27 \times 10^{-23}$~eV~\cite{LIGOScientific:2021sio}. Measurements of the GW wave train also constrain Lorentz violation~\cite{Ellis:2016rrr,LIGOScientific:2021sio}, which could modify the GW dispersion relation in an energy-dependent manner. Both these constraints can be strengthened in the future by detectors operating at lower frequencies and observing the binary infall for longer periods~\cite{Ellis:2020lxl}. 

Another interesting class of measurements is of the final stages of the merger during which the ``dumbbell”-shaped pair of contiguous BHs relaxes to a Kerr BH through the emission of higher-order GW modes~\cite{Fairhurst:2023beb}. These measurements probe the strong-gravity regime and could distinguish models of Exotic Composite Objects (ECOs)~\cite{Banks:2023eym}. Unfortunately, to my knowledge no powerful test of quantum gravity using these measurements has yet been proposed.

Could the BHs whose mergers have been observed by LIGO and Virgo have originated from other astrophysical mechanisms other than stellar collapse, such as earlier mergers? Or could BSM physics be at work to modify standard stellar collapse models, see, e.g.,~\cite{Croon:2020oga}?
Or could these objects be BHs that originated from BSM physics, e.g., as primordial black holes (PBHs)~\cite{Carr:2023tpt}?

\section{Mergers of neutron stars - kilonovae}

In addition to mergers of BHs, LIGO and Virgo have also observed mergers of BHs with neutron stars (NSs) and NS-NS mergers~\cite{LIGOScientific:2017vwq}. The latter are of interest for nuclear physics and QCD, as they can be used to constrain NS masses, radii and equations of state. In principle, they can also be used to constrain BSM physics. For example, if there are lumps of dark matter trapped inside NSs they could modify the expected pattern of GW emissions during NS-NS mergers, adding additional features in the spectrum~\cite{Ellis:2017jgp}. The first and most celebrated NS-NS merger observed so far has been GW170817, which was detected (near-)simultaneously as the gamma-ray burst (GRB) 170817A by the Fermi and INTEGRAL satellites~\cite{LIGOScientific:2017zic}. The coincidence between the GW and gamma-ray observations verified the long-conjectured origins of short GRBs and implied that the velocities of light and GWs are the same to within $\sim 10^{-15}$. 

\section{Do we owe our existence to GWs?}

Dozens of telescopes followed up the observation of GW170817/GRB 170817A, observing its decaying light curve in different frequency bands~\cite{LIGOScientific:2017ync}. These observations verified that NS-NS mergers (known to astronomers as kilonovae) are copious sources of heavy elements that are produced by rapid neutron capture in dense matter (the astrophysical $r$-process). These elements are expected to include iodine and bromine, which play essential roles in human biology~\cite{Nussey,NRC,mccall2014bromine}. They have not been observed directly in the GW170817 kilonova, but there are reports to have observed in this~\cite{Hotokezaka+2023} and another kilonova~\cite{levan2023heavy} an unstable isotope of tellurium, which is adjacent to iodine in the periodic table, and expected to be produced at a similar rate.

\begin{figure}
    \centering
    \includegraphics[height=0.5\textwidth]{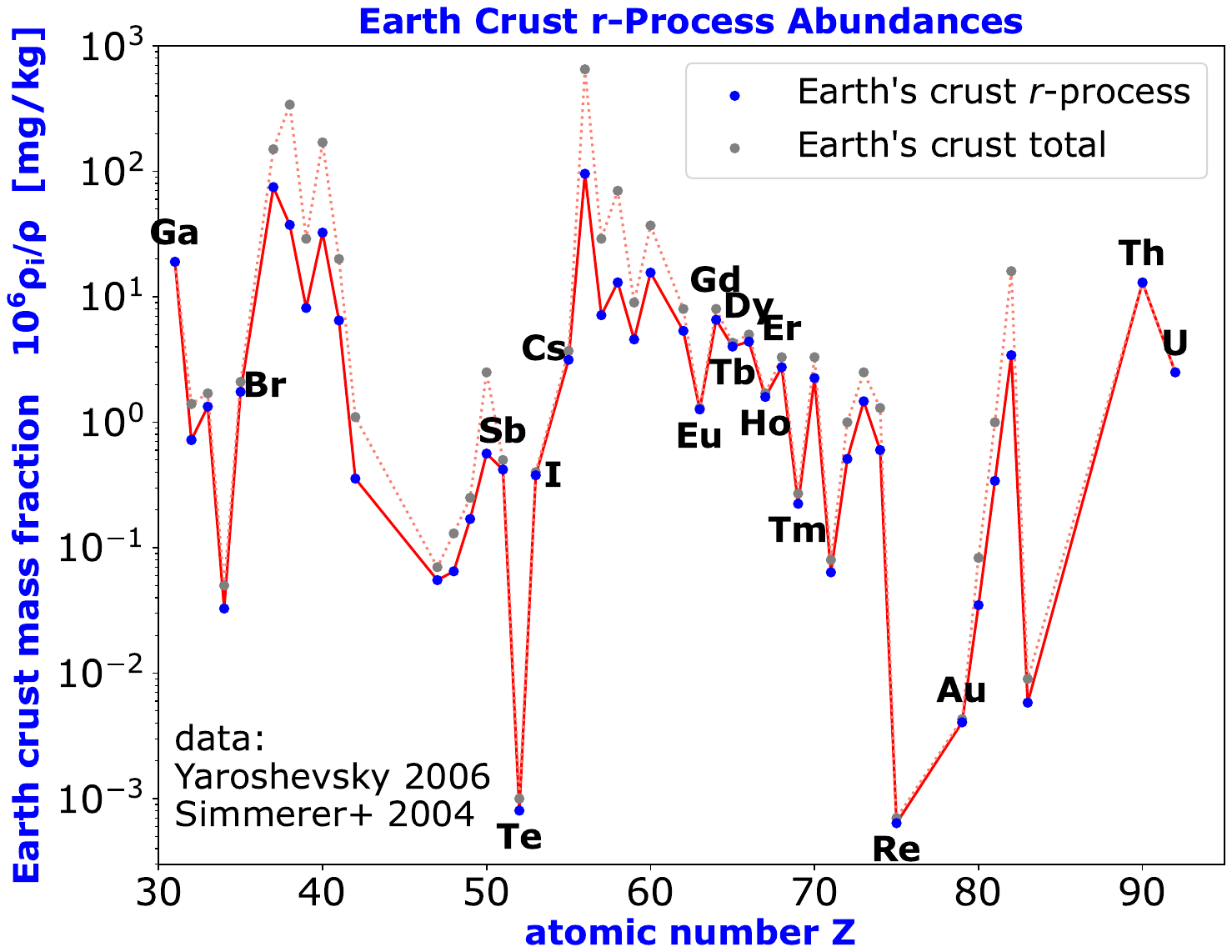}
    \caption{A comparison of the heavy element abundances measured in the Earth's crust (dotted line) with the
    $r$-process portion (solid line), as calculated in~\cite{Ellis:2024uvj}. Elements for which the calculated 
    $r$-process fractions exceed 75\% of the crust abundance are labelled. These include 
    bromine and iodine, which are essential for human biology. 
    }
    \label{fig:rprocess}
\end{figure}

These observations raise the provocative question whether we owe our existence to GWs~\cite{Ellis:2024uvj}? As seen in Fig.~\ref{fig:rprocess}, calculations indicate that the majority of heavy elements in the Earth's crust, including iodine and bromine, were produced by the $r$-process. As discussed above, kilonovae are certainly $r$-process sites, though certain types of supernova are also possible sites. If our iodine and bromine originate from kilonovae, what caused them? Kilonovae occur because of GWs, whose emission caused these NS-NS mergers, as indicated first by the Hulse-Taylor pulsar~\cite{Hulse:1974eb} and now confirmed by the LIGO-Virgo measurements of GW170817~\cite{LIGOScientific:2017vwq}.

\section{Supermassive black holes}

Astrophysicists tell us that, in addition to the stellar mass black holes whose mergers were discovered by LIGO and Virgo, the centres of galaxies are populated by supermassive BHs (SMBHs) weighing millions, even billions of solar masses. Measurements of the orbits of stars near the centre of our own galaxy confirm that it contains a BH weighing some $4\times 10^6$ solar masses~\cite{Ghez:2008ms,Gillessen:2008qv}, and the Event Horizon Telescope has imaged photon rings around this~\cite{EventHorizonTelescope:2022wkp} and the SMBH weighing some $6\times 10^9$ solar masses at the core of the M87 galaxy~\cite{EventHorizonTelescope:2019dse}. LISA~\cite{Barausse:2020rsu,Colpi:2024xhw}, Taiji~\cite{Ruan:2018tsw} and TianQin~\cite{TianQin:2015yph} are laser interferometer space missions that target GWs emitted by SMBHs when they merge or absorb small BHs. With these aims in mind, they are optimised to detect GWs with frequencies $\sim 10^{-2}$~Hz, significantly below the frequencies $\sim 10^2$ Hz where LIGO and Virgo have optimal sensitivities. 

The prevalence of SMBHs raises the pressing question: how have they been assembled~\cite{rees1978quasars} (see the discussion around Fig.~2),~\cite{Woods:2018lty}? Perhaps they were seeded by Primordial BHs? Or perhaps they were put together by a hierarchical process starting from BHs weighing $\sim 10^3$ solar masses that were formed by the collapses of a first generation of massive (Pop~III) stars? Or perhaps somewhat heavier BHs were formed inside protogalaxies that subsequently merged to form those around us today? Or perhaps the SMBHs were formed directly following the mergers of protogalaxies? A promising way to distinguish between different scenarios is to search for intermediate mass BHs (IMBHs) and their possible mergers at different redshifts. There is some astrophysical evidence for IMBHs with masses $\sim 10^4 - 10^5$ solar masses~\cite{Chilingarian:2018acs}. As seen in Fig.~\ref{fig:strains}, their mergers would emit GWs in the frequency range $10^{-2} - 10^2$~Hz that is not covered by existing and planned laser interferometers. There is a gap in the frequency coverage of GW detectors that needs to be filled.

\section{Atom interferometers}

As seen in Fig.~\ref{fig:strains}, the GW detection gap could be filled by atom interferometer (AI) detectors~\cite{Dimopoulos:2008sv,Buchmueller:2023nll}. These are based on the same underlying principles as laser interferometers, but use laser pulses to manipulate clouds of cold atoms, instead of optical beam-splitters and mirrors, as illustrated in Fig.~\ref{fig:Interferometers}. AI projects are underway in the US (MAGIS)~\cite{MAGIS-100:2021etm}, the UK (AION)~\cite{Badurina:2019hst}, France (MIGA)~\cite{Canuel:2017rrp}, Germany (VLBAI)~\cite{schlippert2020matter} and China (ZAIGA)~\cite{Zhan:2019quq}. I am a member of the AION collaboration, which plans to build a sequence of increasingly ambitious detectors, staring with a 10-m interferometer to be built in the basement of the Oxford physics department, to be followed by detectors on the 100-m and 1-km scales, and eventually a space mission called AEDGE~\cite{AEDGE:2019nxb}. 

\begin{figure}[h]
\centering 
\includegraphics[width=0.45\textwidth]{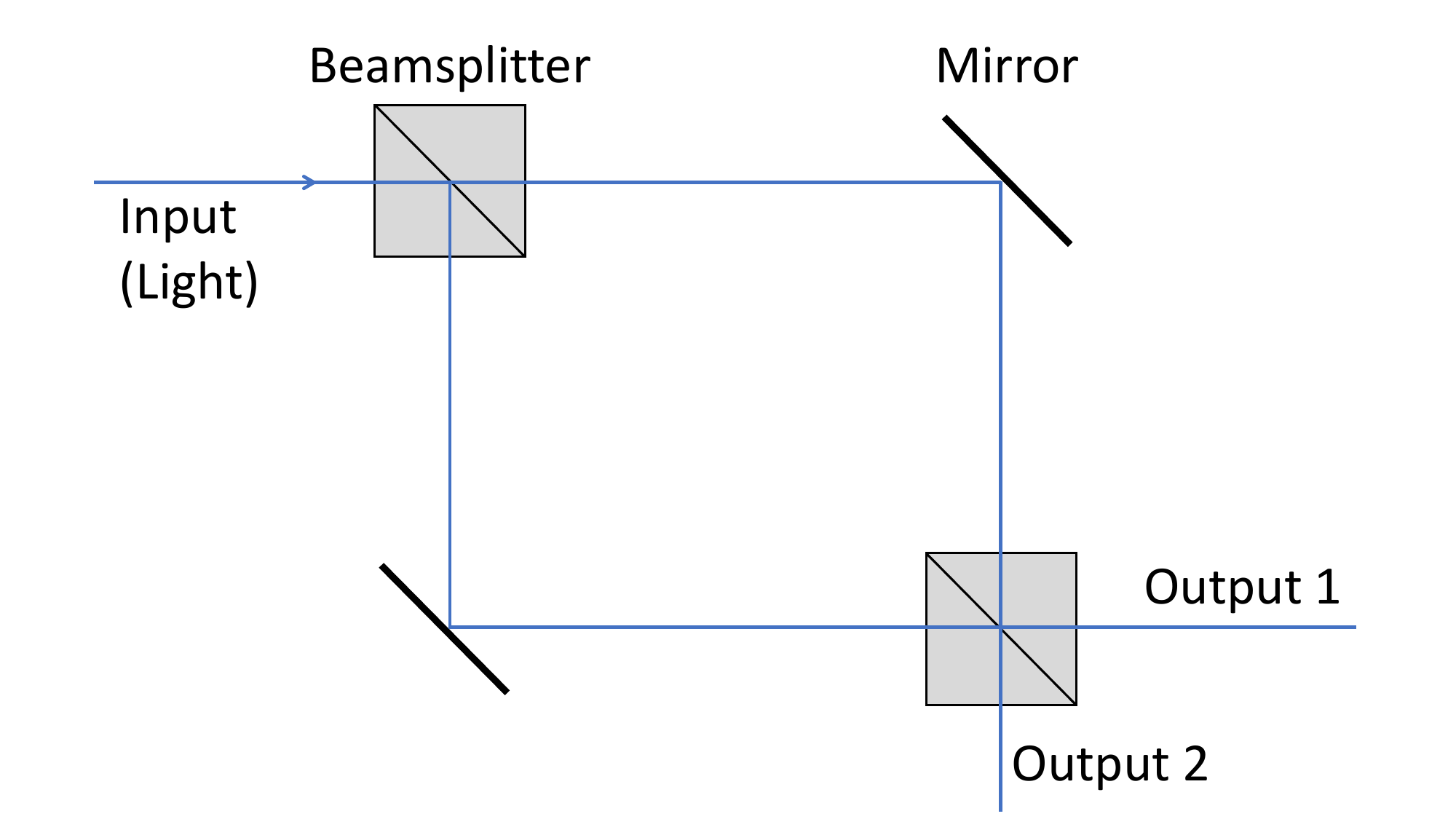}
\includegraphics[width=0.45\textwidth]{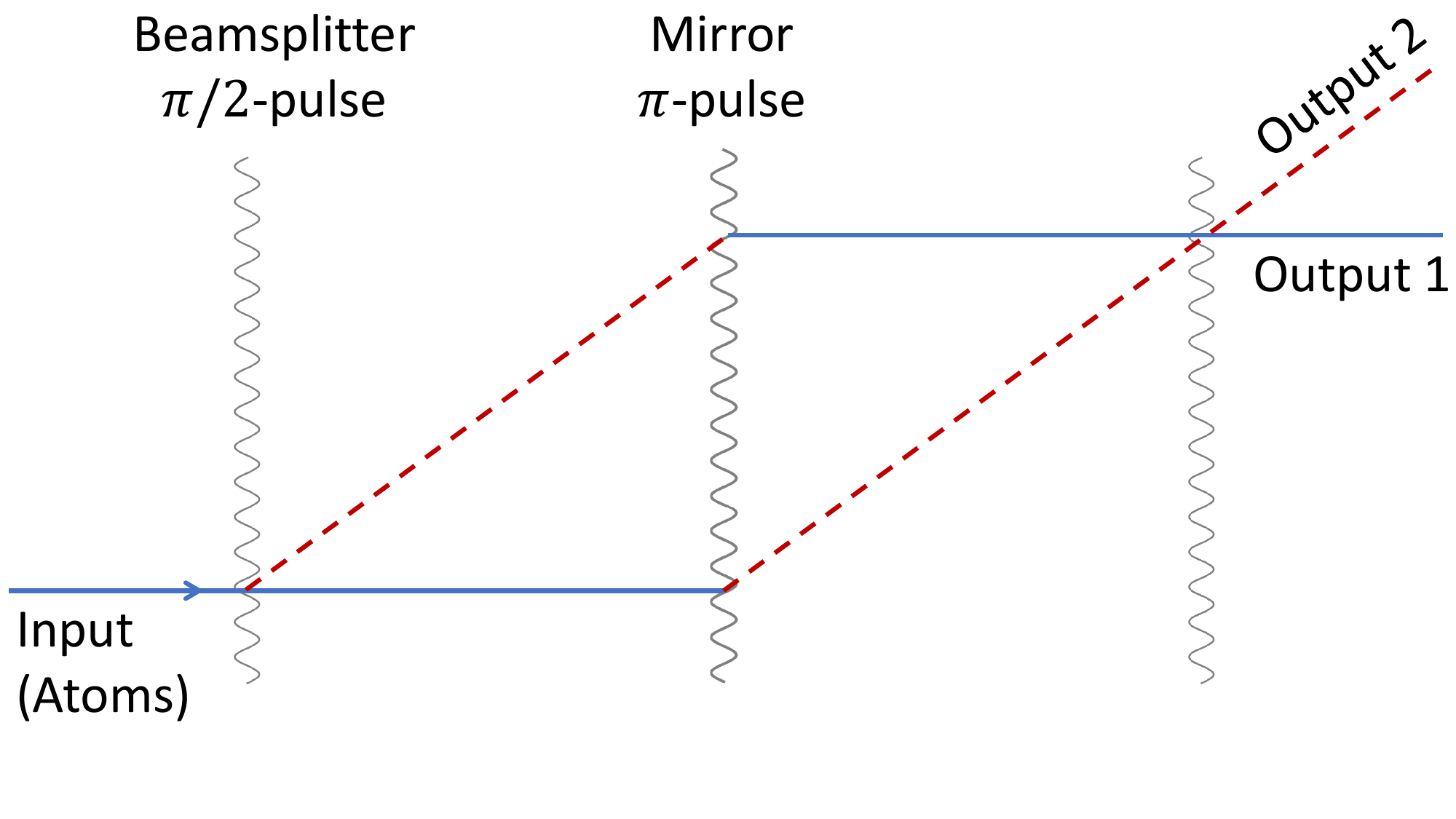}

\caption{Left: Outline of the principle of a laser interferometer. The laser beams are represented by solid blue lines.
Right: Outline of an analogous atom interferometer. Here the solid blue lines represent atoms in the ground state, $|{g}$, the dashed red lines represent atoms in the excited state, $|{e}$, and laser pulses are represented by wavy lines.}
\label{fig:Interferometers}
\end{figure}

As seen in Fig.~\ref{fig:strains}, the larger AI detectors would be able to detect IMBH mergers~\cite{Badurina:2021rgt}, complementing LISA and LIGO-Virgo. They would also have important synergies with these detectors. For example, LISA could observe the early inspiral stages of IMBH mergers whose final stages would by measured by an AI detector, while an AI detector could measure the early stages of a lower-mass BH merger whose final stages could be measured by LIGO and Virgo. The AI measurements could be used to predict the directions, redshifts and timings of such mergers~\cite{Ellis:2020lxl}, facilitating multimessenger observations that could be used to identify the environments in which such mergers occur.

AI experiments also have many applications to searches for BSM physics. For example, they would be able to improve significantly on the sensitivities of LIGO and Virgo to a possible graviton mass and certain forms of Lorentz violation, by virtue of their capabilities to observe long trains of inspiral GWs at lower frequencies~\cite{Ellis:2020lxl}. They would also be able to probe whether merging objects are actually BHs or possibly ECOs~\cite{Banks:2023eym}, by studying the waveforms of their mergers. In parallel, AI experiments can probe for coherent waves of ultralight bosonic matter interacting with atomic constituents~\cite{Geraci:2016fva,Arvanitaki:2016fyj}. These can also be constrained by measuring IMBH spins, which would be suppressed by the superradiance of light bosons~\cite{brito2020superradiance}.

\begin{figure}[h]
\centering 
\includegraphics[width=0.45\textwidth]{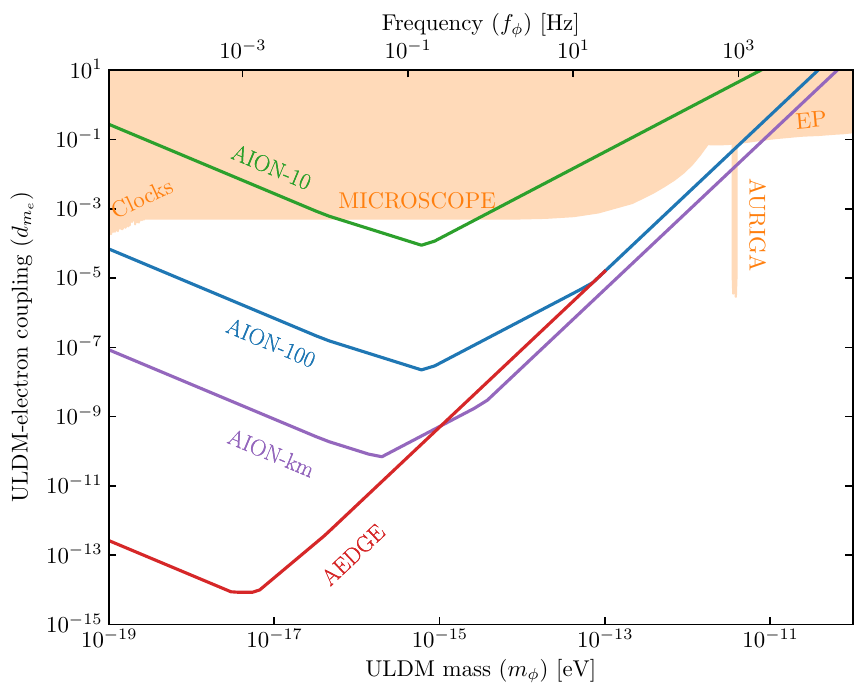}
\includegraphics[width=0.45\textwidth]{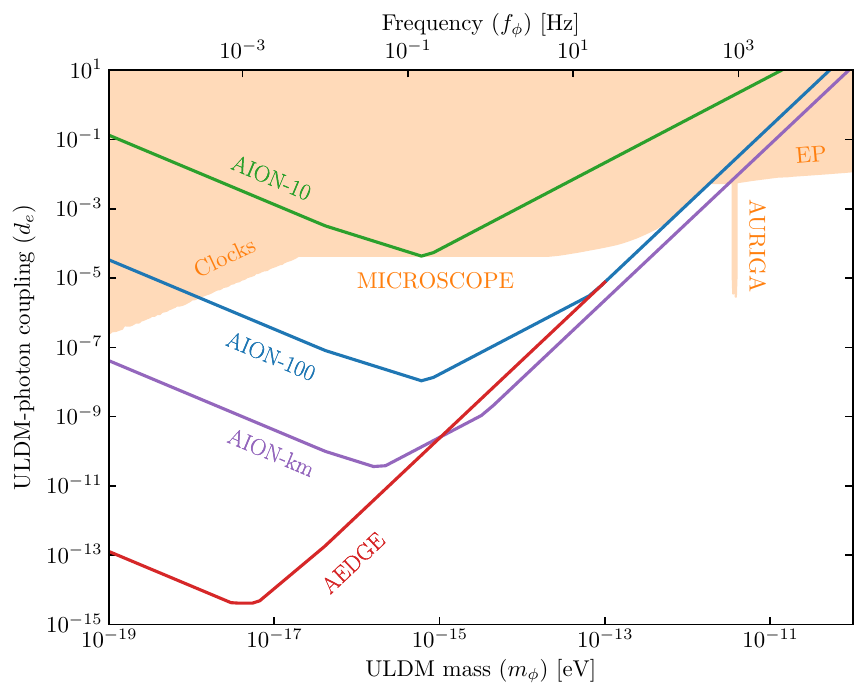}
\caption{Sensitivity projections for linear ULDM couplings to electrons (left panel) and photons (right
panel) with parameters given in~\cite{Badurina:2019hst}, neglecting gravity gradient noise. The green (blue) (purple) and red lines are for AION-10 (-100) (-km)
and AEDGE, respectively. The shaded orange region is excluded by the existing constraints from searches for
violations of the equivalence principle by the MICROSCOPE experiment and with torsion balances, atomic clocks, and the AURIGA experiment, as described in~\cite{Badurina:2019hst}.}
\label{fig:Interferometers}
\end{figure}

\section{Discovery of GWs by PTAs}

A new frontier in GW studies was initiated in 2023 by the discovery by several Pulsar Timing Array (PTA) collaborations of a stochastic background of GWs in the nanoHz frequency range~\cite{NANOGrav:2023gor,EPTA:2023fyk,Reardon:2023gzh,Xu:2023wog}. The PTAs had previously found evidence for a common noise source causing correlated delays in the arrival times of pulsar signals~\cite{Antoniadis:2022pcn}: the new element was the observation of an angular correlation in the noise that is characteristic of GWs, as first predicted by Hellings and Downs~\cite{Hellings:1983fr}. The default astrophysical interpretation of this signal is GW emissions by a population of SMBH binaries - precursors to the biggest bangs since the Big Bang - which predicts a characteristic frequency dependence, as first calculated by Phinney~\cite{Phinney:2001di}.

\begin{figure}
    \centering
    \includegraphics[width=0.8\textwidth]{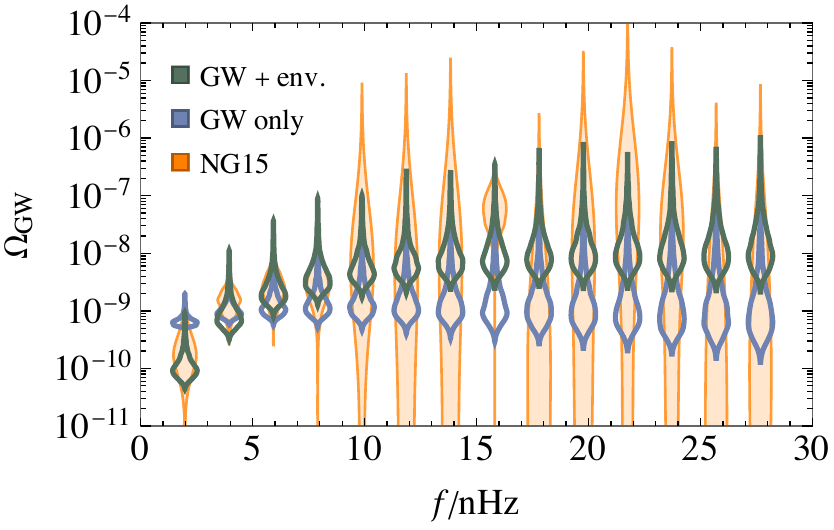}
    \caption{The NANOGrav 15-year data on the energy density of GWs, $\Omega_{GW}$, (orange), compared with the best fit assuming that BH binaries lose energy by GW emission only (blue), and the best fit allowing also for energy loss by environmental effects (green)~\cite{Ellis:2023dgf}.}
    \label{fig:OmegaGWfits}
\end{figure}

However, the frequency dependence of the PTA data does not agree very well with this interpretation, particularly when statistical sampling effects are taken into account~\cite{NANOGrav:2023hvm,Ellis:2023dgf}. We analysed the possibility that the binaries lose energy by interactions with their environments, as well as by GW emission. There are several mechanisms for environmental energy loss, including interactions with stars and dynamical friction or viscous drag due to the surrounding medium~\cite{Kelley:2016gse}. Quantifying these possible effects is a complex issue, so we adopted a simple phenomenological parameterisation of environmental energy loss and showed that this could improve the description of the PTA data~\cite{Ellis:2023dgf}, as seen in Fig~\ref{fig:OmegaGWfits}.

However, the PTA data may also be explained by BSM physics~\cite{NANOGrav:2023hvm,EPTA:2023xxk} and related cosmological phenomena, and some of these alternative models provide even better fits to the PTA data~\cite{Ellis:2023oxs}~\footnote{Apologies for not referring to the many papers on this subject, a large sampling of which are cited in~\cite{Ellis:2023oxs}.}. One such possibility is provided by GWs generated at first or second order during cosmological inflation, and scenarios of this type have been proposed within specific BSM models~\cite{Basilakos:2023xof,Basilakos:2023jvp}. Another possibility is that the background GWs were generated by a network of cosmic strings~\cite{Ellis:2023tsl}, mainly when string loops collapse. Alternatively, the nanoHz GWs might have been generated by a first-order cosmological phase transition, or by domain walls. Either of these scenarios would have needed a dynamical scale in the GeV range~\cite{Ellis:2023oxs}, and would have to have occurred in some hidden sector. Another possible source is an “audible” axion model. Although these BSM models fit the data better than astrophysical scenarios, the improvement is not (yet?) highly significant, and it would be premature to throw your BSM hat in the air.

\section{Distinguishing interpretations of PTA GW signals}

There are ways in which future data could help distinguish between astrophysical and BSM scenarios. For example, astrophysical scenarios predict fluctuations between the signals in different frequency bins as well as anisotropies and polarisation effects~\cite{Ellis:2023oxs}, whereas the BSM scenarios typically predict relatively smooth, isotropic and unpolarised signals. These differences arise from the fact that the astrophysical signal is dominated by relatively few nearby sources, which could in principle be distinguished.

\begin{figure}
    \centering
    \includegraphics[width=0.9\textwidth]{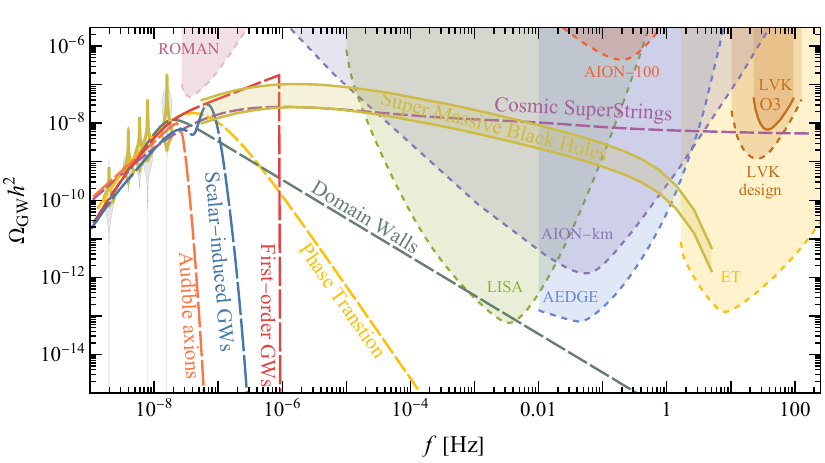}
    \caption{Comparison of the extensions of fits to NANOGrav 15-year data to higher frequencies, indicating the prospective sensitivities of LIGO-Virgo-KAGRA and planned and proposed future detectors~\cite{Ellis:2023oxs}. The green band extends the fit of the SMBH binary scenario allowing for environmental effects from the PTA range to higher frequencies, and shows the mean GW energy density spectrum from SMBH binaries heavier than $10^3$ solar masses for BH merger probabilities in the range 0.25 - 1. Individual SMBH binaries are expected to be measurable at frequencies above the PTA range.}
    \label{fig:BFplotsbig}
\end{figure}

\begin{figure}
    \centering
    \includegraphics[width=0.6\textwidth]{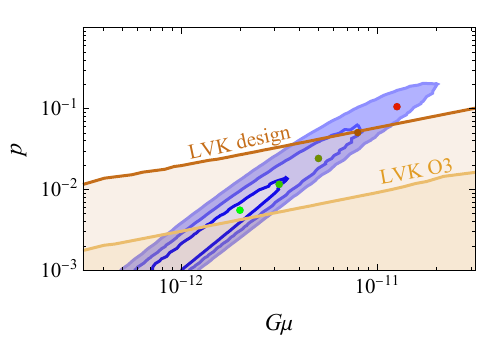}
    \caption{The blue contours correspond to the 68, 95 and 99\% CL ranges for string fits to the NANOGrav 15-year data assuming standard cosmological expansion~\cite{Ellis:2023tsl}. The coloured dots correspond to the curves with specific values of the string tension $G \mu$ and intercommutation probability $p$. The orange regions show the present and projected LIGO/Virgo/KAGRA sensitivities.}
    \label{fig:stringLVK}
\end{figure}

Also, significant differences between the predictions of the different scenarios appear at frequencies above the nanoHz range explored by the PTAs. As seen in Fig.~\ref{fig:BFplotsbig}, most of the BSM scenarios predict spectra of GWs that fall off above this range, and would not be detectable by LISA and higher-frequency experiments~\cite{Ellis:2023oxs}. An exception is provided by the cosmic string scenario, which predicts a frequency spectrum extending into the LIGO-Virgo-KAGRA range~\cite{Ellis:2023tsl}. Indeed, this scenario is already constrained  by existing LIGO-Virgo-KAGRA data, and could be probed quite thoroughly by future LIGO-Virgo-KAGRA data~\cite{Ellis:2023oxs}, as seen in Fig.~\ref{fig:stringLVK}

On the other hand, one can extrapolate the GW predictions of SMBH binary scenarios to higher frequencies using an extended Press-Schechter formalism for the mergers of galactic haloes and assuming a constant probability for those mergers to yield merging SMBH binaries~\cite{Ellis:2023owy,Ellis:2023dgf}. Under these assumptions, LISA should be able to detect individual mergers of BHs weighing $10^6 - 10^9$ solar masses, whereas AI experiments would be mergers of BHs weighing $10^3 - 10^6$ solar masses. The combination of LISA and AI experiments could probe the different scenarios for the (hierarchical?) assembly of SMBHs~\cite{Ellis:2023iyb}.

\section{Quo vadis PTAs?}

Astrophysics or fundamental physics? Have the PTAs detected GW precursors of the biggest bangs since the Big Bang, or have they detected BSM physics? The jury is still out, but preliminary conclusions are that SMBH binaries driven by GWs alone are disfavoured as the source of the PTA GW signal, that SMBH binaries driven by a combination of GWs and environmental effects fit the PTA data better, and even better fits can be obtained within some cosmological BSM models~\cite{Ellis:2023oxs}. However, these preferences are not definitive, and await confirmation with future data from the PTAs and other GW experiments. The good news is that discrimination between the models should be possible with future measurements, including bin-to-bin fluctuations in the frequency spectra, anisotropies and polarisation of the GWs.  Looking ahead, experiments at higher frequencies including LISA and atom interferometers should be able to distinguish astrophysical and BSM scenarios~\cite{Ellis:2023oxs} and explore how SMBHs were seeded and assembled~\cite{Ellis:2023iyb}.

Time and more data will tell whether the PTAs have indeed detected GW precursors of the biggest bangs since the Big Bang, which would excite primarily astrophysicists, or whether they have detected BSM physics, which would be more exciting for particle physicists.

\section*{Funding}

This work was supported in part by the United Kingdom STFC Grants ST/X000753/1 and ST/T00679X/1.

\reftitle{References}


\bibliography{Gravitational_Waves_Echoes}


\PublishersNote{}
\end{document}